\documentclass[preprint]{elsarticle}

\usepackage{amsmath,euscript,amssymb,epsfig,amsfonts,latexsym}
\usepackage{graphicx,color}

\newcommand{\be}[1]{\begin{equation}\label{#1}}
\newcommand{\ee}{\end{equation}}
\newcommand{\ba}[1]{\begin{eqnarray}\label{#1}}
\newcommand{\ea}{\end{eqnarray}}
\newcommand{\rf}[1]{(\ref{#1})}
\newcommand{\nn}{\nonumber}

\usepackage{hyperref}

\begin{document}

\begin{frontmatter}

\title{Mass density vs. energy density\\ at cosmological scales }

\author[a]{Maxim Eingorn\corref{mycorrespondingauthor}}
\cortext[mycorrespondingauthor]{Corresponding author}
\ead{maxim.eingorn@gmail.com}

\author[b,c,d] {Ezgi Yilmaz}
\author[d]{A.~Emrah Y\"{u}kselci}
\author[b,c,e]{Alexander Zhuk}

\address[a]{Department of Mathematics and Physics, North Carolina Central University, \\1801 Fayetteville St., Durham, North Carolina 27707, U.S.A.}
\address[b]{Center for Advanced Systems Understanding,\\ Untermarkt 20, 02826 G\"{o}rlitz, Germany}
\address[c]{Helmholtz-Zentrum Dresden-Rossendorf,\\ Bautzner Landstra\ss e 400, 01328 Dresden, Germany}
\address[d]{Department of Physics, Istanbul Technical University,\\ 34469 Maslak, Istanbul, T\"urkiye}
\address[e]{Astronomical Observatory, Odessa I.I. Mechnikov National University, \\ Dvoryanskaya St. 2, Odessa 65082, Ukraine}

\begin{abstract}
In the presence of the gravitational field, the energy density of matter no longer coincides with its mass density. A discrepancy exists, of course, also between the associated power spectra. Within the $\Lambda$CDM model, we derive a formula that relates the power spectrum of the energy density to that of the mass density and test it with the help of N-body simulations run in comoving boxes of 2.816 Gpc/$h$. The results confirm the validity of the derived formula and simultaneously show that the power spectra diverge significantly from one another at large cosmological scales.	
\end{abstract}


\end{frontmatter}

\

\section{Introduction}
\label{sec:introduction}

Upcoming galaxy surveys such as \cite{euclid,lsst} are expected to probe the distribution of large-scale inhomogeneities at unprecedented scales and with remarkable precision. When studying the large-scale structure via N-body simulations, especially at a wide range of scales including gigaparsec-scale regions, it must be taken into account that relativistic effects play an important role in large domains and hence, the corresponding numerical code should be solving the complete system of general relativistic field equations.

Recently, the all-scale cosmological perturbation theory has been formulated in a relativistic approach, which employs linearized field equations \cite{Eingorn1,fluids2}
that can be solved analytically for metric perturbations. It has been shown in this scheme that the first-order expression for the gravitational potential has a Yukawa-type form, and the gravitational force starts to decay exponentially at scales of 2-3 Gpc \cite{MaxEzgi} unlike what is expected of its Newtonian counterpart. Employing the publicly available relativistic code \textit{gevolution} \cite{gevNat, gevolution}, this so-called \textit{screening} of gravity at large cosmological scales has been studied also via N-body simulations in \cite{delta-rho}, in terms of the behaviour of the power spectrum of the mass density contrast. As expected, it has been revealed that growth of matter overdensities is suppressed beyond the time-dependent characteristic cutoff scale, i.e. the \textit{screening length}. 

In the presence of the gravitational field, the energy density of matter (which corresponds to the mixed 00-component of the energy-momentum tensor) differs from its mass density \cite{Landau}, and it is surely nontrivial to obtain a formula connecting the power spectra of the two quantities. From the point of view of screening formalism, the quantity relevant to matter fluctuations is the mass density contrast \cite{Eingorn1}. The analytical derivation of such a formula is provided in the present work for the $\Lambda$CDM model. It is then tested via studying the behaviour of the power spectra of both quantities, extracted from N-body simulations carried out in boxes of 2.816 Gpc/$h$ in comoving size.

The paper is structured as follows: in Section 2, we derive a formula that relates the power spectrum of the energy density to that of the mass density in the $\Lambda$CDM model. In Section 3, we verify this formula based on simulation outputs. Results are briefly summarized in concluding Section 4.

\

\section{Relation between the power spectra of energy and mass densities}
\label{sec:gevolution}

In the $\Lambda$CDM model, the perturbed Friedmann-Lema\^{\i}tre-Robertson-Walker metric
in the conformal Newtonian gauge reads \cite{Mukhanov2,Rubakov}
\be{1}
ds^2 = a^2 \left[ (1+2\Phi)d\eta^2 - (1-2\Phi)\delta_{ij} dx^i dx^j \right]\, ,\quad i,j=1,2,3\, ,
\ee
where, within the scope of our research, we consider only first-order scalar perturbations $\Phi$ (with $|\Phi|\ll 1$), representing the gravitational potential.
$a(\eta)$ is the scale factor (depending on the conformal time $\eta$) and $x^i$ denote the comoving coordinates. 

For cold dark matter (CDM), fluctuations of the energy density in the first-order are described as \cite{Eingorn1,Hubble} 
\be{2}
\delta\varepsilon\equiv \varepsilon - \bar\varepsilon =  \frac{c^2}{a^3} \delta\rho + \frac{3c^2}{a^3} \overline\rho \Phi\, ,
\ee
where
\be{3}
\delta\rho\equiv\rho-\bar\rho\,
\ee
is the comoving mass density fluctuation. The average value of the energy density $\bar\varepsilon$ is related to that of the comoving mass density as $\bar\varepsilon = \bar\rho c^2/a^3$. On the other hand, from the Friedmann equation we get 
\be{4}
\bar\varepsilon =\frac{1}{\kappa}\left(\frac{3{\mathcal{H}^2}}{a^2}-\Lambda\right),
\ee
where $\mathcal{H}\equiv(da/d\eta)/a$ is the conformal Hubble parameter and $\kappa\equiv 8\pi G/c^4$, with $c$ standing for the speed of light and $G$ for the Newtonian gravitational constant.

As follows from Eq.~\rf{2}, for non-vanishing gravitational field (the potential $\Phi$), $\delta\varepsilon$ deviates from the definition of mass density fluctuations, which reads $\delta\rho/a^3$.  Within the weak field expansion scheme of the cosmic screening approach, and based on the effective description of peculiar velocities (as worked out explicitly in \cite{MaxEzgi}), the gravitational potential $\Phi$ satisfies the Helmholtz-type equation  
\be{5}
\bigtriangleup \Phi -\frac{a^2}{\lambda_{\rm{eff}}^2}\Phi =\frac{\kappa c^2}{2a}\delta\rho\, ,
\ee
which follows from the linearized time-time component of field equations below:
\be{lin00}
\triangle \Phi-3\mathcal{H}\left(\Phi'+\mathcal{H}\Phi\right)=\frac{1}{2}\kappa a^2 \delta\varepsilon\,.
\ee
Here $\bigtriangleup\equiv \delta^{ij}\partial_{i}\partial_j$ (with $\partial_{i}\equiv \partial/\partial x^{i}$) is the Laplace operator while the prime denotes the derivative with respect to the conformal time, and  
\be{6}
\lambda_{\rm{eff}}(\eta)=
\left(\frac{a\mathcal{H}}{3}\int_0^a\frac{d\tilde a}{ \mathcal{H}^3}\right)^{1/2}\, 
\ee
determines the characteristic length of the Yukawa-type cutoff for gravitational interactions. For $\Lambda$CDM cosmology and based on the Planck data reported in \cite{planck2018}, its current value is found to be $\lambda_{{\rm{eff}},0}=2.57$ Gpc.

Eq.~\rf{2} shows the relationship between fluctuations in the energy density and mass density. Now, in order to formulate the relationship between the power spectra of these quantities, we resort to the definition \cite{gevNat} 
\be{7}
4\pi k^3\langle \widehat{X}({\bf{k}},z)\, \widehat{X}^*(\tilde{\bf{k}},z)\rangle =(2\pi)^3\delta_{\rm D} ({\bf k}-\tilde{\bf k})P_{X}(k,z)\, 
\ee
that applies to some variable  $X({\bf{r}},z)$ at redshift $z$, where $\delta_{\rm D} ({\bf k}-\tilde{\bf k})$ stands for the Dirac delta function. The hat denotes the Fourier transform of the respective quantity. The dimensionless power spectrum $P_{X}(k,z)$ \cite{0112551} is the Fourier transform 
\be{8}
P_{X}(k,z) = 4\pi k^3\int \frac{d^3{\bf{r}}}{(2\pi)^3}\xi_{X} (r,z)\exp(i{\bf{k}} {\bf{r}})
\ee
of the two-point correlation function
\be{9}
\xi_{X}(r,z)=\frac{1}{V}\int d{\bf{x}} X({\bf{x}},z)X ({\bf{x}}+{\bf{r}},z)\,,
\ee
defined in the volume $V$. It is worth noting that an alternative definition of the power spectrum is also often used in the literature, that reads $\tilde P_X(k,z)=\left(2\pi^2/k^3\right)P_X(k,z)$ in units of $({\rm Mpc}/h)^3$ \cite{Rubakov}, which, unlike the behaviour presented in our work in the upcoming figures, decreases for larger modes after the turnover at $k=k_{\rm eq}$ indicating the size of horizon at matter-radiation equality.  Due to statistical homogeneity and isotropy, $\xi_X$ depends on $r\equiv|{\bf{r}|}$, and $P_X$ depends on $k\equiv|{\bf{k}}|$ \cite{0112551}. From this point on we drop the redshift dependence to avoid cluttered notation.

Based on Eq.~\rf{2}, the correlation function for energy density fluctuations can be written in the form
\ba{10}
\xi_{\delta\varepsilon}(x)&=&\frac{1}{V}\int d{\bf{r}} \delta\varepsilon({\bf{r}})\delta\varepsilon ({\bf{r}}+{\bf{x}})\nn\\
&=&\frac{1}{V}\int d{\bf{r}}\left[\frac{c^2}{a^3} \delta\rho({\bf r}) + \frac{3c^2}{a^3} \overline\rho \Phi({\bf r})\right] \left[\frac{c^2}{a^3} \delta\rho({\bf r}+{\bf x}) + \frac{3c^2}{a^3} \overline\rho \Phi({\bf r}+{\bf x})\right]\nn \\
&=& \left(\frac{c^2}{a^3}\right)^2\xi_{\delta\rho}(x)+ \left(\frac{3c^2}{a^3}\overline\rho\right)^2\xi_{\Phi}(x)\,\nn\\
&+&3\overline\rho\left(\frac{c^2}{a^3}\right)^2\frac{1}{V}\left[\int d{\bf{r}}\delta\rho({\bf r})\Phi({\bf r}+{\bf x}) +\int d{\bf{r}} \Phi({\bf r})\delta\rho({\bf r}+{\bf x})\right]\,.
\ea
To calculate the integrals in the last row, we note that
\ba{11}
\bigtriangleup_{\bf x}\xi_{\Phi}(x) &=&\frac{1}{V}\int d{\bf r}\Phi({\bf r})\bigtriangleup_{\bf x}\Phi({\bf r}+{\bf x})\nn\\
&=&\frac{1}{V}\int d{\bf r}\Phi({\bf r})\left[\frac{a^2}{\lambda_{\rm{eff}}^2}\Phi({\bf r}+{\bf x})+ \frac{\kappa c^2}{2a}\delta\rho ({\bf r}+{\bf x}) \right]\, ,
\ea
where we make use of use Eq.~\rf{5}. Consequently, we get
\be{12}
\frac{1}{V}\int d{\bf{r}} \Phi({\bf r})\delta\rho({\bf r}+{\bf x})
=\left(\frac{\kappa c^2}{2a}\right)^{-1}\left[ \bigtriangleup_{\bf x}\xi_{\Phi}(x)- \frac{a^2}{\lambda_{\rm{eff}}^2}\xi_{\Phi}(x)\right]
\, .\ee
Meanwhile, 
\be{13}
\int d{\bf{r}} \Phi({\bf r})\delta\rho({\bf r}+{\bf x})
=\int d{\bf{r}} \Phi({\bf r}-{\bf x})\delta\rho({\bf r})
=\int d{\bf{r}} \Phi({\bf r}+{\bf x})\delta\rho({\bf r})\, ,
\ee
where in the last equality, we exploited the invariance of the right-hand side (RHS) of Eq.~\rf{12} with respect to the sign of ${\bf x}$. Since the left-hand side (LHS) must also be invariant under the same transformation, the correlation function for the energy density fluctuation reads
\be{14}
\xi_{\delta\varepsilon}(x)=\frac{c^4}{a^6}\xi_{\delta\rho}(x)  +9{\bar\varepsilon}^2\xi_{\Phi}(x) + \frac{12}{\kappa} \bar\varepsilon\left[\frac{1}{a^2}\bigtriangleup_{\bf x}\xi_{\Phi}(x) - \frac{1}{\lambda_{\rm{eff}}^2}\xi_{\Phi}(x)\right]\, .
\ee
Finally, for the power spectra, we obtain the formula
\be{15}
P_{\delta\varepsilon}(k)= \frac{c^4}{a^6}P_{\delta\rho}(k)+3\bar\varepsilon\left[3\bar\varepsilon-\frac{4}{\kappa}\left(\frac{k^2}{a^2}+\frac{1}{\lambda_{\rm{eff}}^2}\right)\right]P_{\Phi}(k)\, ,
\ee
where we have taken into account that the Fourier transform of $\bigtriangleup_{\bf x}\xi_{\Phi}(x)$ gives $-k^2P_{\Phi}(k)$.

\

\section{Validation via N-body simulations}
\label{sec:N-body}



In the following step, we test the validity of the formula in Eq.~\rf{15} employing N-body simulations. It was shown previously in \cite{vs1} that the power spectra of scalar (and vector) perturbations agree to high accuracy when simulated according to the screening approach versus the default formulation of the relativistic code $gevolution$, despite the differences in how the weak field expansion is performed within two schemes. For the current study, we have simulated $P_{\delta\varepsilon}$ on the LHS of Eq.~\rf{15} using the code \cite{gevNat,gevolution} without any modifications. Namely, the particles were evolved using Eq.~(8) of \cite{vs1} and the scalar perturbation was updated via Eq.~\rf{extra1} below. The resulting behaviour of the spectrum has also provided a reference point, given that \textit{gevolution} agrees with the perturbation theory to great extent at large scales investigated here, as verified in \cite{gevNat}. The power spectra $P_{\delta\rho}$ and $P_{\Phi}$ on the RHS of Eq.~\rf{15}, however, were obtained in a second run,
based on the set of equations of the cosmic screening approach: the $\Phi$-dependent term in the full expression for $\delta\varepsilon=\delta T^0_0=T^0_0-\overline{T_0^0}$ (where the energy-momentum tensor element $T_0^0$ is given by (3.7) of \cite{gevolution}, taking into account (3.11) and (3.13) therein) was moved to the LHS of the corresponding field equation in \textit{gevolution}, that is, (2.9) in \cite{gevolution}, and the remaining term $\propto\delta\rho$ was left as the single first-order source on the RHS upon removing the extra terms proportional to squared momenta, negligible for nonrelativistic bodies with the comoving mass density  
$\rho=\sum_n m_n\delta({\bf r}-{\bf r}_n)$ in our configuration. All second-order contributions were also removed from the LHS to match the weak field expansion employed in the screening approach (based on the reasoning discussed in detail in \cite{vs1}). In brief, the evolution equation for the scalar mode in the code \cite{gevolution}, that is (for $c=1$) 
\ba{extra1}
\left(1+4\Phi\right)\triangle\Phi-3\mathcal{H}\Phi'+3\mathcal{H}^2\left(\chi-\Phi\right)+\frac{3}{2}\delta^{ij}\partial_{i}\Phi\partial_{j}\Phi=-4\pi G a^2\left(T^0_0-\overline{T^0_0}\right)\,,\nn\\
T^0_0=-\frac{1}{a^4}\sum_n\delta\left({\bf r- \bf r}_n\right)\sqrt{q_n^2+m_n^2a^2}\left(1+3\Phi+\frac{q_n^2}{q_n^2+m_n^2a^2}\Phi\right)\,,\quad\quad
\ea
was reduced to
\ba{extra2}
&&\triangle\Phi-3\mathcal{H}\Phi'-3\Phi\left(\mathcal{H}^2+4\pi G\overline{\rho}/a\right)\,\nn\\
&=&-4\pi G a^2\left(-\sum_n m_n\delta({\bf r}-{\bf r}_n)/a^3-\overline{T^0_0}\right)\,.
\ea
Particles' momenta were evolved using $d{\bf q}_n/d\eta=-m_n a\nabla \Phi$. The function $\chi$ denotes the difference between scalar modes in the perturbed metric due to anisotropic stress. It is a second-order quantity in the screening approach in the absence of relativistic species, and thus was dropped together with terms quadratic in the gravitational potential.

Introducing the notations
\ba{16}
P^{\rm g}_{\delta}&\equiv& P_{\delta\varepsilon/{\bar\varepsilon}}=\frac{1}{\bar\varepsilon^2}P_{\delta\varepsilon}\, ,\\
\label{17} P^{\rm s}_{\delta}&\equiv& P_{\delta\rho/{\bar\rho}}=\frac{1}{\bar\rho^2}P_{\delta\rho}\, ,
\ea 
Eq.~\rf{15} may be reformulated as
\be{18}
P_{\delta}^{\rm g}(k,z)= P_{\delta}^{\rm s} (k,z)+3\left[3-\frac{4}{\kappa{\bar\varepsilon}}\left(\frac{k^2}{a^2}+\frac{1}{\lambda_{\rm{eff}}^2}\right)\right]P_{\Phi}(k,z)\, .
\ee

To test the above expression, we have run two simulations, both in boxes with comoving sizes of 2816 Mpc/$h$ at 1 Mpc/$h$ resolution, as explained above.  For $\Lambda$CDM cosmology, the parameters used were $H_0 = 67.66 \,  {\rm km}\, {\rm s}^{-1} {\rm Mpc}^{-1}$, $\Omega_{\rm M}h^2 = \Omega_{\rm b}h^2+\Omega_{\rm c}h^2=0.14175$, where $h=0.6766$,
and $\Omega_{\Lambda}=1-\Omega_{\rm M}$ \cite{planck2018}. Baryons and cold dark matter have been treated on equal footing and the scale factor at the present time was set to $a_0=1$.

\begin{figure*}
	\resizebox{1.\textwidth}{!}{\includegraphics{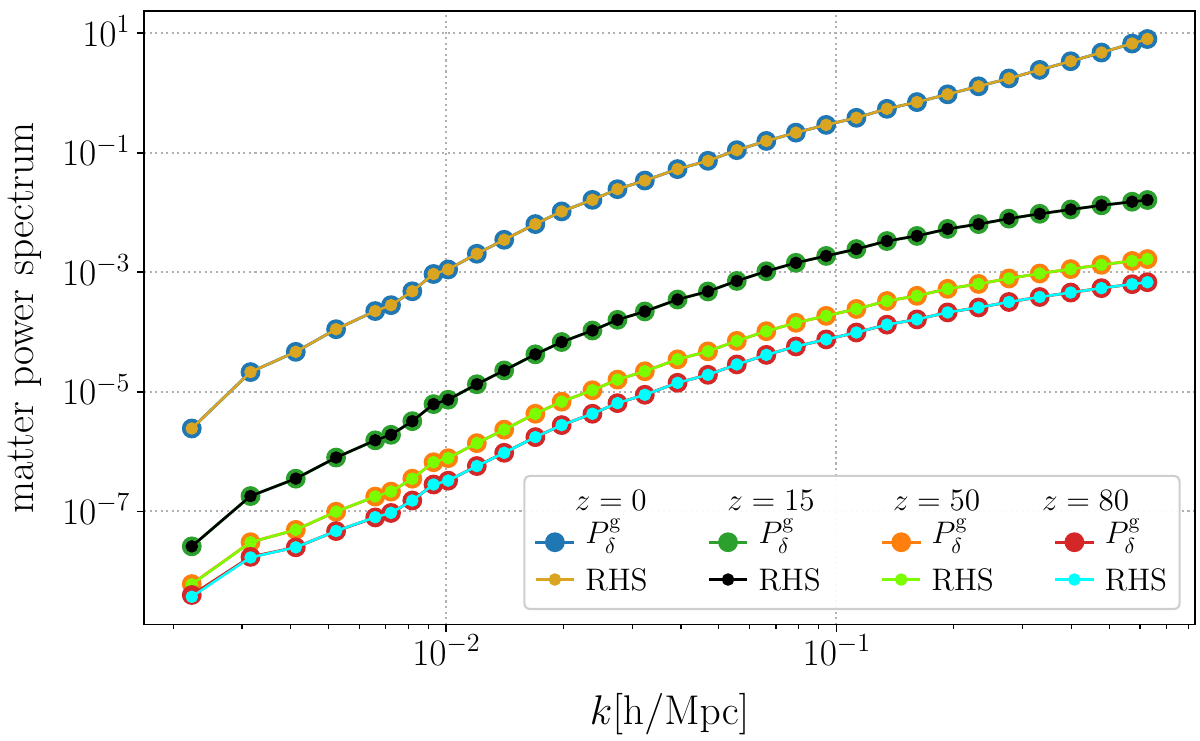}}
	\caption{Four pairs of curves for dimensionless matter power spectra from the left- ($P^{\rm g}_{\delta}$) and right-hand sides of Eq.~\rf{18} at redshifts $z=80, 50, 15$ and $z=0$ from bottom to top.}
	\label{fig:1}
\end{figure*}

\begin{figure*}
	\resizebox{1.\textwidth}{!}{\includegraphics{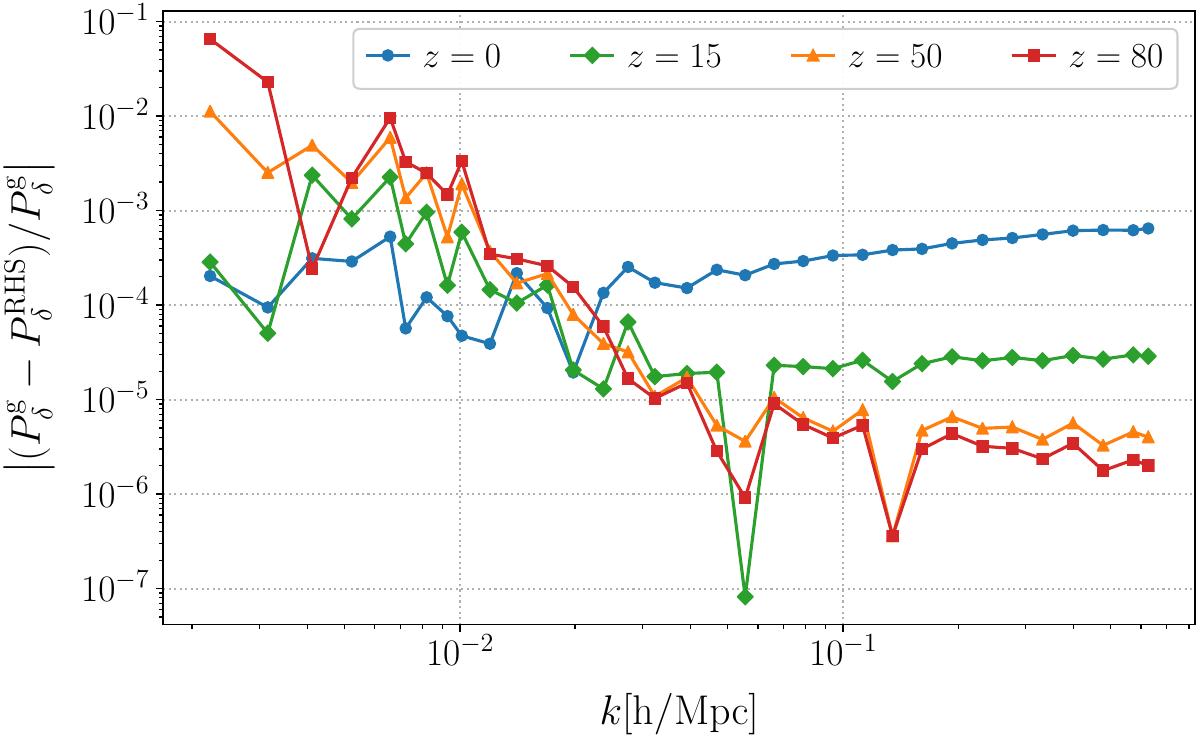}}
	\caption{Relative deviations $\Delta_{\rm rd}$ given by Eq.~\rf{19} for four pairs of curves in Fig~\rf{fig:1}.}
	\label{fig:2}
\end{figure*}

\begin{figure*}
	\resizebox{1.\textwidth}{!}{\includegraphics{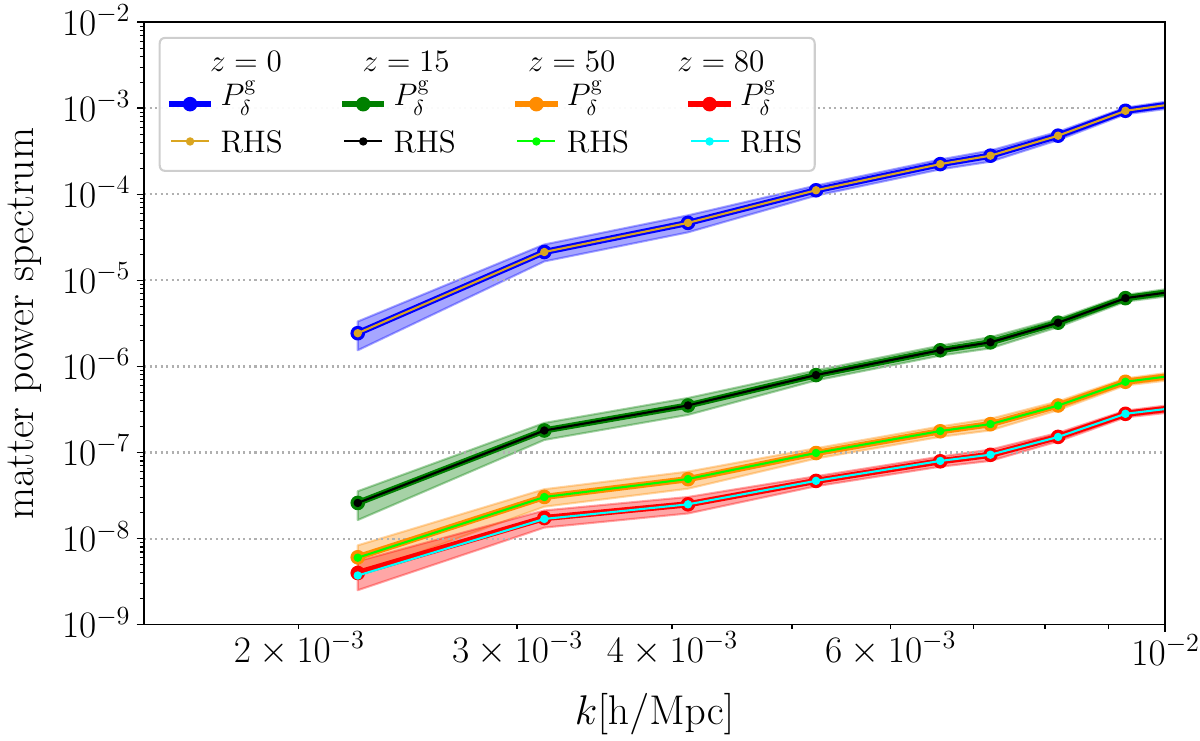}}
	\caption{Standard deviations of dimensionless power spectra  $P^{\mathrm{\rm g}}_{\delta}$  at redshifts $z=80, 50, 15$ and $z=0$ from bottom to top. All curves that correspond to the RHS of Eq.~\rf{18} remain within the respective shaded regions.}
	\label{fig:3}
\end{figure*}

Four pairs of curves are depicted in Fig.~\ref{fig:1} corresponding to redshifts $z=80, 50, 15$ and $z=0$. In each pair, one curve corresponds to the LHS of Eq.~\rf{18} (i.e. $P_{\delta}^{\rm g}$), and the second one to the full expression on the RHS.
To estimate the difference between the left- and right-hand sides of \rf{18} quantitatively, we introduce the relative deviation
\be{19}
\Delta_{\rm rd} \equiv \left|\frac{P^{\mathrm{g}}_{\delta}-P^{\mathrm{RHS}}_{\delta}}{P^{\mathrm{g}}_{\delta}}\right|
\ee
for each pair. Fig.~\ref{fig:2} demonstrates that relative deviations reach their maximum values at small momenta $k$ (i.e. large scales) and large redshifts. Nevertheless, the curves which correspond to the RHS of \rf{18} are within the standard deviation of $P^{\mathrm{g}}_{\delta}$ (see Fig.~\ref{fig:3}). So large values of $\Delta_{\rm rd}$ are mostly associated with the inaccuracy of calculations at the larger scales in the simulation, intrinsically represented with relatively less data points in the k-space of the box used. We stress that the standard deviation indicated by the shaded regions in the graph is the deviation about the mean, represented by the data points in the figure. As expected, each data point is generated using a number of $k$ values, and the shaded regions basically show how the value of the power spectrum may deviate from the mean.

\begin{figure*}
	\resizebox{\textwidth}{!}{\includegraphics{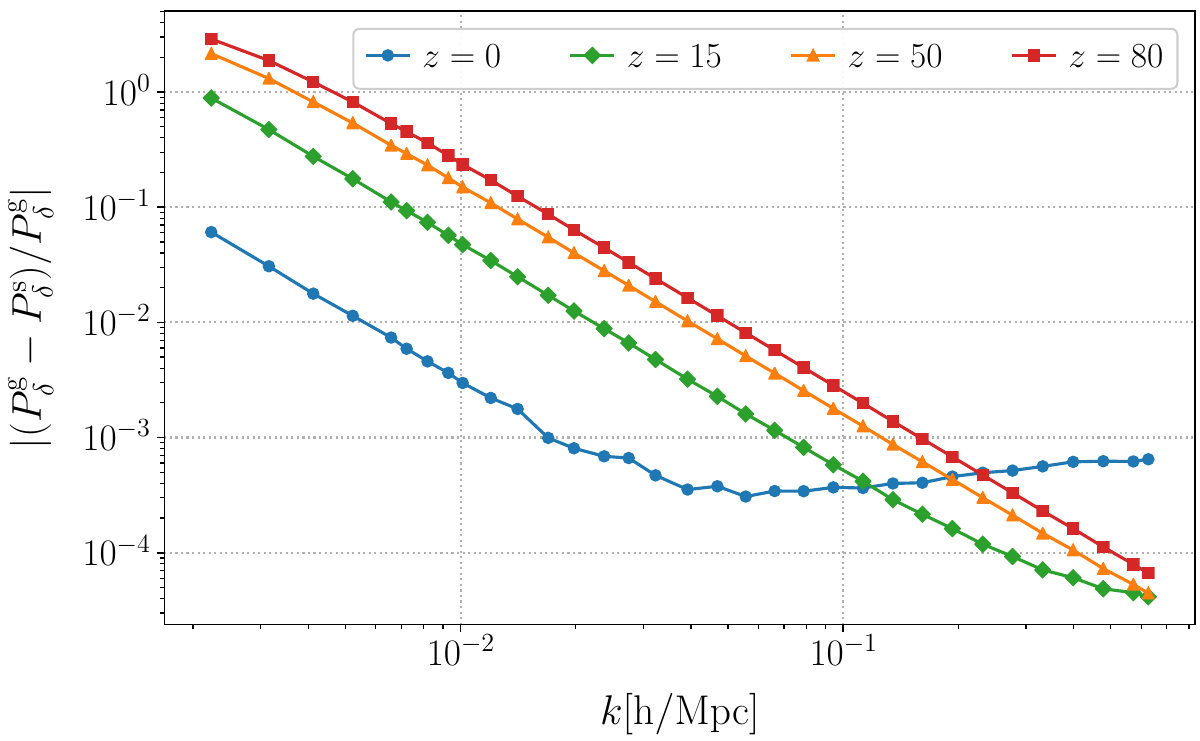}}
	\caption{Relative deviation for the dimensionless power spectra $P^{\mathrm{g}}_{\delta}$ and $P^{\mathrm{s}}_{\delta}$ at redshifts $z=80, 50, 15$ and $z=0$. This figure demonstrates the significant divergence at small momenta $k$ and large redshifts.}
	\label{fig:4}
\end{figure*}

It is, of course, of interest to compare the power spectrum of energy density contrast $P^{\mathrm{g}}_{\delta}$ to that of the mass density contrast $P^{\mathrm{s}}_{\delta}$, too. The associated relative deviation curves are presented in Fig.~\ref{fig:4}, which indicate a larger splitting for small momenta and large $z$. This is because, at large cosmological scales, the second term on the RHS of Eq.~\rf{18}, $\propto P_\Phi$, becomes non-negligible with respect to the power spectrum for the mass density contrast $P_\delta^s$.

\ 

\section{Conclusion}
\label{sec:conclusion}

In the present paper we have considered the power spectra of energy and mass density fluctuations for the $\Lambda$CDM model.  The energy density of matter does not coincide with the mass density in the presence of the gravitational field, and the respective power spectra differ also. In attempt to quantify the discrepancy, we have first derived the analytical expression for the relationship between two power spectra, and then tested it with the help of N-body simulations carried out in boxes of 2.816 Gpc/$h$ comoving sizes. We have calculated the power spectrum of the energy density contrast using \textit{gevolution} \cite{gevolution}.  To calculate the power spectrum of the mass density contrast, we have modified the code to recover the corresponding equations in the cosmic screening approach \cite{delta-rho,vs1}.  Numerical simulations have confirmed that the derived formula works well. We have also shown that the power spectra of the energy density contrast and the mass density contrast diverge significantly at large cosmological scales where relativistic effects are essential.

Number counts on our past light cone are one of the relevant observables when studying the large-scale structure, which are subject to relativistic effects emerging due to the perturbed geometry of  spacetime as well as the peculiar motion of the sources such as galaxies. Detailed derivation of separate contributions are available in literature within the framework of the linear perturbation theory, suitable for studying large scales with small density contrasts. As discussed in \cite{cr,cb} (and references therein), an accurate formula takes into account (energy) density perturbations of matter with a galaxy bias, redshift space distortion due to peculiar velocities, lensing, and the additional influence of large-scale relativistic effects on the propagation of light rays. Our results zoom into the first contribution here, and indicate a splitting in the expression for energy density perturbations in the adopted gauge, prior to the investigation of the observed distribution of point sources.


\

\section*{Acknowledgments}

{\noindent This work was partially supported by the Center for Advanced Systems Understanding (CASUS) which is financed by Germany's Federal Ministry of Education and Research (BMBF) and by the Saxon state government out of the State budget approved by the Saxon State Parliament. Computing resources used in this work were provided by the National Center for High Performance Computing of T\"urkiye (UHeM) under grant number 4007162019.}

\section*{Data Availability}

\sloppy

{\noindent The datasets generated in this work are publicly available in the Rossendorf Data Repository (RODARE) and may be accessed via the link \newblock\href{https://rodare.hzdr.de/record/2469}{{https://rodare.hzdr.de/record/2469}}}.

\

\end{document}